\documentclass[aip,floats,superscriptaddress,onecolumn,preprint]{revtex4-1}
\usepackage{amssymb}
\usepackage{amsmath}\usepackage{graphicx}

\usepackage{color}
\graphicspath{{../../Fig_final.d/}}

\begin{document}
\title{Charge carrier mobility in hybrid halide perovskites}

\author{Carlo Motta}
\email[]{mottac@tcd.ie}
\affiliation{School of Physics, AMBER and CRANN Institute, Trinity College, Dublin 2, Ireland}
\author{Fedwa El-Mellouhi}
\email[]{felmellouhi@qf.org.qa}
\affiliation{Qatar Environment and Energy Research Institute, Doha, Qatar}
\author{Stefano Sanvito}
\affiliation{School of Physics, AMBER and CRANN Institute, Trinity College, Dublin 2, Ireland}

\begin{abstract}
The charge transport properties of hybrid halide perovskites are investigated with a combination of density functional theory
including van der Waals interaction and the Boltzmann theory for diffusive transport in the relaxation time approximation. We 
find the mobility of electrons to be in the range 5-10~cm$^2$V$^{-1}$s$^{-1}$ and that for holes within 
1-5~cm$^2$V$^{-1}$s$^{-1}$, where the variations depend on the crystal structure investigated and the level of doping. 
Such results, in good agreement with recent experiments, set the relaxation time to about 1~ps, which is the time-scale
for the molecular rotation at room temperature. For the room temperature tetragonal phase we explore two possible 
orientations of the organic cations and find that the mobility has a significant asymmetry depending on the direction of the 
current with respect to the molecular axis. This is due mostly to the way the PbI$_3$ octahedral symmetry is broken. 
Interestingly we find that substituting I with Cl has minor effects on the mobilities. Our analysis suggests that the carrier mobility 
is probably not a key factor in determining the high solar-harvesting efficiency of this class of materials.
\end{abstract}

\date{\today}

\maketitle


Hybrid halide perovskites have made a breakthrough in the field of organic solar cells~\cite{Green2014}. Their unique energy-harvesting 
efficiency, combined with the low manufacturing costs position them as an ideal materials class to focus research. The conversion of sunlight 
into electrical power has recently surpassed the outstanding efficiency of 15\% for both mesoporous metal-oxide scaffolds and in planar 
heterojunction architectures~\cite{Burschka:2013,Liu:2013,Liu_Kelly_2013} and the latest studies report a value exceeding
 18\%~\cite{C4EE00942H}. A unique property of these materials is the ability to both 
act as light-harvesting medium and as charge carrier transporter. However, despite the intense research carried out in the past two years, 
some questions remain open regarding the nature of the material's working principles. For instance, it is still under debate whether the 
photo-generated charges have an excitonic or a free-carrier character, with results pointing towards contrasting
conclusions~\cite{doi:10.1021.jz5005285,ADMA:ADMA201305172,doi:10.1021jz500858a}.
Also, the origin of the high efficiencies of perovskite-based solar cell devices still needs to be unraveled, and it is not completely clear how 
the mobility of the active layer influences the overall performance.

Savenije et al.\cite{doi:10.1021jz500858a} have recently performed microwave photo-conductance and photo-luminescence experiments, 
measuring a mobility of 6.2~cm$^{2}$/Vs at 300~K. They found a band-like dependence of the mobility with temperature with a slope of $T^{-1.6}$. 
At the same time, using transient THz spectroscopy, Wehrenfennig and coworkers have shown that CH$_{3}$NH$_{3}$PbI$_{3}$ exhibits 
long charge-carrier diffusion lengths, exceeding 1~$\mu$m, and a high-frequency mobility of 8~cm$^{2}$/Vs, an indeed remarkable result 
for a solution-processed material~\cite{ADMA:ADMA201305172}. They inferred that the low bimolecular recombination rate arises from a 
spatial separation of electrons and holes in the system. Finally, early dielectric measurements~\cite{Poglitsch1987} suggest a picosecond 
relaxation process at room temperature.

To our knowledge, to date no theoretical calculations have addressed the problem of evaluating the conductivity and charge mobility in hybrid 
perovskites. Motivated by this gap in the knowledge and by the intriguing r\^ole that the charge transport is expected to play, here we present 
a first-principles analysis of the transport properties of organolead halide perovskites. In particular we consider the methyl-ammonium 
lead-iodide perovskite, CH$_{3}$NH$_{3}$PBI$_{3}$, which is the archetype of this class of materials. The cubic, tetragonal, and orthorhombic 
phases are explored. We show that, depending on the phase of the material and the doping level, the mobility spans a range between 5 and 
12~cm$^{2}$/Vs for holes and 2.5 and 10~cm$^{2}$/Vs for electrons. Furthermore, our results suggest that Cl doping has little impact on the 
transport properties.


The electronic structure of CH$_{3}$NH$_{3}$PBI$_{3}$ has been calculated with the all-electron {\sc fhi-aims} code at the level of the
generalised gradient approximation (GGA) in the Perdew-Burke-Ernzerhof (PBE)~\cite{Perd96} parameterization. Long-range van der 
Waals interactions are included via the Tkatchenko and Scheffler (TS) scheme~\cite{PhysRevLett.102.073005}, which is constructed 
over a GGA and a pairwise dispersive potential. In order to verify the goodness of the bandstructure, additional calculations with the HSE 
\cite{HSE} and HSE06~\cite{HSE06} functional have been performed. Since these give us rather similar mobilities than those obtained with 
GGA, the results are not presented here. The reciprocal space integration was performed over an 8$\times$8$\times$8 Monkhorst-Pack 
grid~\cite{PhysRevB.13.5188} in the case of the cubic cell, and a 6$\times$6$\times$4 one for both the tetragonal and the orthorhombic. 
A pre-constructed high-accuracy all-electron basis set of numerical atomic orbitals was employed, as provided by the {\sc fhi-aims} ``tight'' 
default option. Structural optimization was performed with the Broyden-Fletcher-Goldfarb-Shanno algorithm~\cite{numrec97}, with the crystal 
geometry determined by optimizing both the internal coordinates and the supercell lattice vectors with a tolerance of $10^{-3}$~eV/\AA\ and 
with the constraint of orthogonal cell vectors. 

The charge mobility has been determined by mean of the semiclassical Boltzmann theory within the constant relaxation time approximation, 
as implemented in the {\sc BoltzTrap} code~\cite{g.madsen20064}. The code has been interfaced with {\sc fhi-aims} and uses the 
{\sc fhi-aims}-calculated wave-functions and eigenvalues. A very dense $k$-point sampling of 32$\times$32$\times$32 (32768 $k$-points 
over the full Brillouin zone) has been employed for the cubic cell, while for the tetragonal and orthorhombic cells it was reduced to 
18$\times$18$\times$12.

We now briefly summarize the key steps of the scheme. From the first-principles bandstructure, $\varepsilon_{i,\bf k}$, the $\alpha$ 
component of the group velocity for a charge carrier in the $i$-th band is obtained as
\begin{equation}
 v_{\alpha}(i,{\bf k}) = \frac{1}{\hbar}\frac{\partial\varepsilon_{i,\bf k}}{\partial k_{\alpha}}\:,
\end{equation}
and used to compute the conductivity tensor
\begin{equation}
 \sigma_{\alpha,\beta}(i,{\bf k}) = 
 e^{2}\tau_{i,{\bf k}} v_{\alpha}(i,{\bf k}) v_{\beta}(i,{\bf k})\:,
\end{equation}
where $e$ is the electronic charge and $\tau_{i,{\bf k}}$ is the relaxation time. By integrating $\sigma_{\alpha,\beta}(i,{\bf k})$, one can extract
the conductivity as a function of the temperature, $T$, and the chemical potential, $\nu$,
\begin{equation}\label{eq:3}
 \sigma_{\alpha,\beta}(T,\nu) =
 \sum_{i}\int \frac{d{\bf k}}{8\pi^{3}} \left[ -\frac{\partial f(T,\nu)}{\partial\varepsilon}\right]
 \sigma_{\alpha,\beta}(i,{\bf k}) ,
\end{equation}
where $f$ is the Fermi-Dirac distribution function. Note that $\nu$ is determined by the number of free carriers or, equivalently, by their concentration.
Once the conductivity is known, the charge mobility reads
\begin{equation}\label{eq:4}
 \mu(T,\nu) = \frac{\sigma(T,\nu)}{\rho e}\:,
\end{equation}
where $\rho$ is the free carrier (electron or hole) concentration. While Eq.~(\ref{eq:3}) is in principle exact within the semiclassical theory, it is very 
hard to compute unless some assumptions are taken. Therefore, for practical purposes the transport properties are calculated by introducing two 
approximations: (i) the relaxation time $\tau$ is constant, i.e. it is independent of the temperature, the band index and the chemical potential; (ii) the 
band structure is treated as independent of temperature or doping and, therefore, is fixed independently of the chemical potential. Nonetheless, despite
such approximations, the method has been successful to predict the transport and thermoelectric properties of many 
materials~\cite{C3TA11222E,Wu201262,PhysRevB.83.085204,PhysRevB.85.125209}.

\begin{figure}[h!]
\includegraphics[width=0.6\columnwidth]{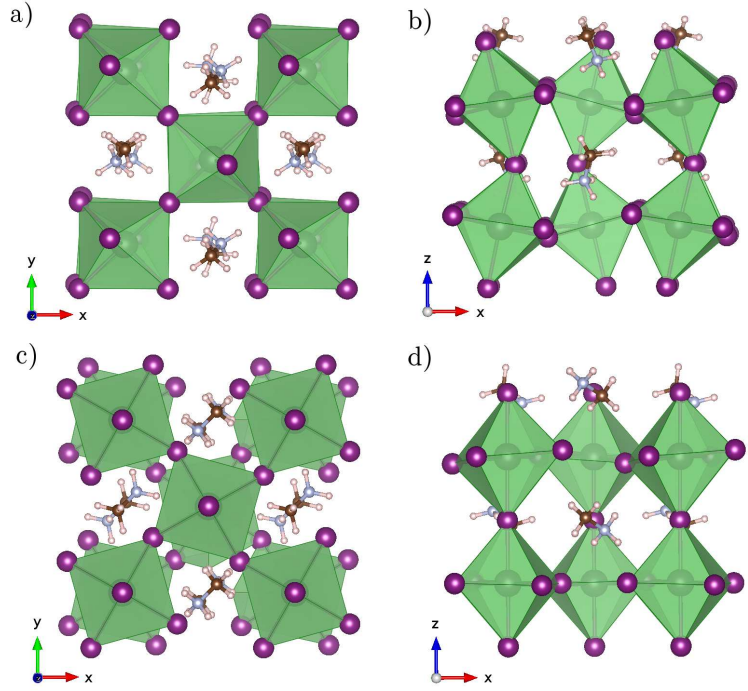} 
\caption{\label{fig:1} Optimized structures for the tetragonal phase of CH$_{3}$NH$_{3}$PbI$_{3}$. Top and side view of \textit{tetra1} are in (a) and (b), 
and for \textit{tetra2} in (c) and (d). The two structures are obtained by initializing the crystal relaxation with the cation oriented along (001) and (110), 
respectively.}
\end{figure}

Here we will mainly focus on the tetragonal phase, the room temperature one, but results will be presented also for the cubic and orthorhombic. 
The geometries for the three phases are optimised by initializing the structures according to the experimental diffraction data~\cite{Stoumpos:2013}. 
In order to determine the impact of the molecular orientation on the stabilization of the inorganic matrix, which we previously discussed in the case of 
the cubic cell~\cite{carlo2014}, we consider two initial orientations for the CH$_{3}$NH$_{3}$ molecule, namely along (001) and (110). The relaxed 
structures are shown in Fig.~\ref{fig:1} and will be referred to as \textit{tetra1} and \textit{tetra2}, respectively. The configurational energy minimum is 
strongly correlated to the orientation of the organic cations, in fact by only twisting the methyl-ammonium orientation two qualitatively different configurations 
emerge. Indeed, our calculations confirm the strong mutual connection between the organic and inorganic subsystems. As a result of the interaction 
between the N and the I species, the (001) orientation induces an alternate tilting of the PbI$_3$ octahedra of $\sim14^{\circ}$ with respect to the (001) 
axis [Fig.~\ref{fig:1}(b)]. In contrast, the (110) orientation induces an alternate tilting of the same amount, $\sim14^{\circ}$, in the (110) plane [Fig.~\ref{fig:1}(c)], 
while no disorder occurs along the (001) axis. The lattice parameter of \textit{tetra1} and \textit{tetra2} are 9.01~\AA\, 8.72~\AA\, 12.36~\AA\ and 
8.71~\AA\, 8.71~\AA\, 12.78~\AA\ respectively, in agreement with experiments~\cite{Stoumpos:2013}. The energy of such two configurations is also 
similar, differing by $20$~meV.

\begin{figure}[h!]
\includegraphics[width=0.6\columnwidth]{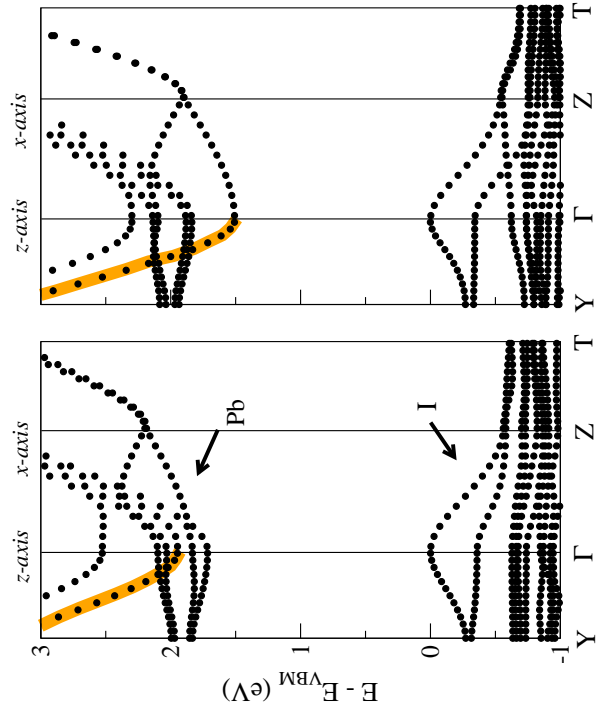} 
\caption{\label{fig:2} Electronic properties of the tetragonal phase of CH$_{3}$NH$_{3}$PbI$_{3}$
                       in the \textit{tetra1} (left) and \textit{tetra2} (right) configurations. }
\end{figure}

The electronic band structures of the two tetragonal phases look also very similar, as depicted in Fig.~\ref{fig:2}. We verified that 
the bottom of the conduction band mainly originates from the $p$ orbitals of Pb, while the top of the valence band is derived from the 
$p$ orbitals of I. The HOMO and LUMO of methyl-ammonium are located approximately at $\pm5$~eV below the valence band maximum (VBM).
The major difference in the bandstructure consists in a reduction of the gap from $1.7$~eV to $1.5$~eV
in the case of \textit{tetra2}, arising from a downshift of the Pb-$5p$ band at the conduction band edge, with a larger separation from the manyfold composed of 
a mixture of Pb-$p$ and I-$p,d$ bands. Accordingly, the band curvature increases around $\Gamma$. A larger conductance close to the conduction band 
edge is thus expected for \textit{tetra2}.

\begin{figure}[h!]
\includegraphics[width=0.6\columnwidth]{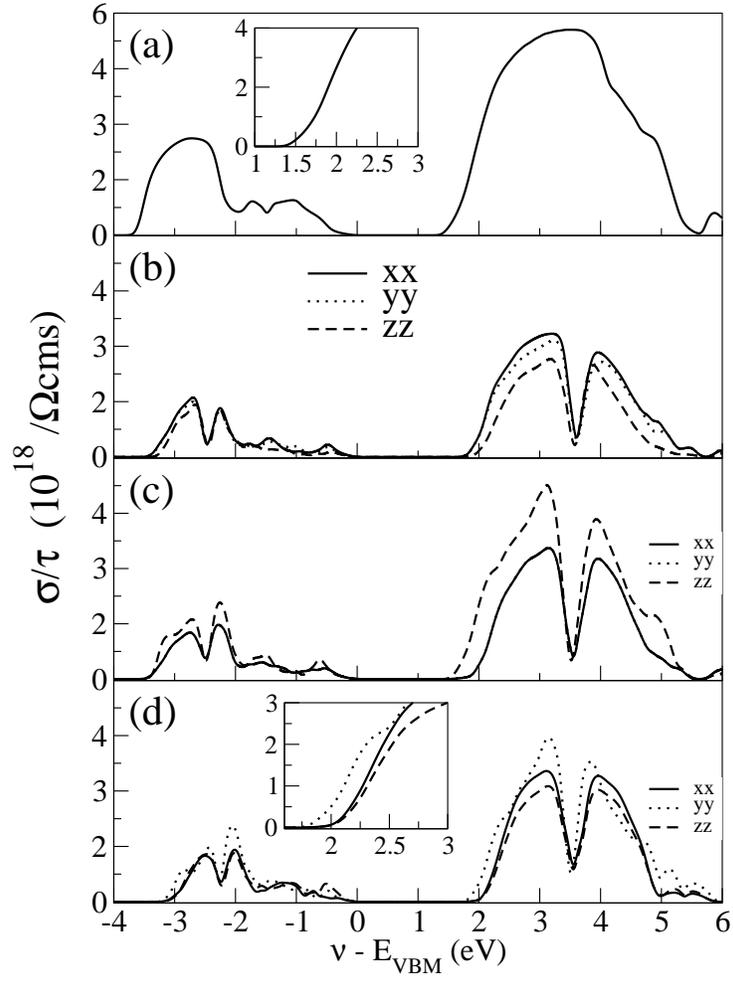} 
\caption{\label{fig:3} Conductivity, $\sigma/\tau$, as a function of the chemical potential for
                       the cubic (a), \textit{tetra1} (b), \textit{tetra2} (c) and orthorhombic (d)
                       phases of CH$_{3}$NH$_{3}$PbI$_{3}$. The solid, dotted and dashed lines 
                       represent the conductivity tensors $\sigma_{xx}$, $\sigma_{yy}$ and $\sigma_{zz}$.}
\end{figure}

The diagonal components of the conductivity tensor, $\sigma_{\alpha}=\sigma_{\alpha,\alpha}$, for the three phases of CH$_{3}$NH$_{3}$PbI$_{3}$ 
are reported in Fig.~\ref{fig:3} as a function of the chemical potential. In the cubic phase the metal sublattice is almost isotropic along the three unit cell 
dimensions. Thus, although the molecular cations may have different orientations, we find that the conductivity itself is isotropic. In fact, the bands around 
the Fermi level, $E_{\rm F}$, originate only from Pb and I. In contrast, due to symmetry lowering the conductivity of the tetragonal and orthorhombic 
structures displays a significant anisotropy. For both \textit{tetra1} and \textit{tetra2}, $\sigma_{x}$ and  $\sigma_{y}$ assume very similar values. 
Interestingly, in \textit{tetra1} $\sigma_{x,y}$ is larger than $\sigma_{z}$, while for \textit{tetra2} we see the opposite behaviour. This can be explained 
in terms of the distortion of the octahedra. In \textit{tetra1}, the octahedra tilting occurs along the $z$ axis, therefore the overlap of the I-$p$ orbital reduces 
and the transport along that direction is suppressed. In contrast, the disorder in \textit{tetra2} is mainly in the $x,y$ plane, while along $z$ the I-Pb-I bonds 
are aligned causing $\sigma_{z}$ to be larger. For the orthorhombic geometry, being it qualitatively similar to \textit{tetra2}, the same considerations
can be made.
%
We point out here that our calculations are performed without the inclusion of 
spin-orbit coupling (SOC). Indeed, SOC is known to play a role in lead-iodide materials, mainly due to the presence of Pb. 
However, after comparing the bandstructures with those obtained with SOC, we noticed that the major effect 
is a renormalization of the energy gap. 
For completeness, we computed the effective masses along two high-symmetry
directions around the $\Gamma$ point with and without the SOC, as reported in Tab.~\ref{tab:1}. 
In agreement with previous calculations~\cite{Umari:2014,Giorgi:2013}, it turns out that the effect of SOC is sizeable, 
yet not dramatic.

\begin{figure}[h!]
\includegraphics[width=0.6\columnwidth]{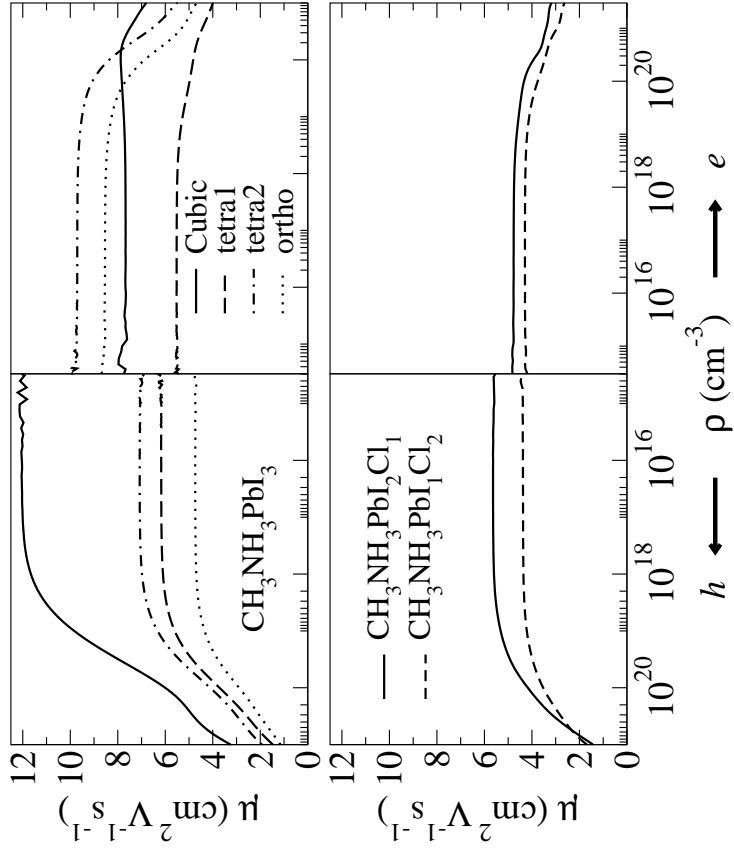} 
\caption{\label{fig:4} Average hole and electron mobility for the different phases of CH$_{3}$NH$_{3}$PbI$_{3}$ (upper panel)
                      as a function of the charge concentration, $\rho$. The average is taken over the three diagonal components
                      of the mobility tensor. In the lower panel the same quantity is plotted for Cl-doped perovskites.}
\end{figure}

We can now turn our discussion to the mobility, which has been calculated from Eq.~(\ref{eq:4}) by considering $\tau = 1$~ps, as determined 
experimentally in Ref.~\citenum{Poglitsch1987}. 
The agreement that we find between our calculated mobilities and those recently measured 
experimentally suggests that such estimate is indeed a valid one. The specially averaged holes and electrons mobilities, $\mu$, are plotted in 
Fig.~\ref{fig:4} as a function of the charge carrier concentration, $\rho$, a quantity that is experimentally hard to determine with certainty, as it 
depends on the device processing, and on the presence of defects and impurities. It has been often argued that in most of the measurements 
$\rho$ is actually very low, corresponding to a nearly intrinsic semiconductor~\cite{Stoumpos:2013}. In any case, our calculations show a generally 
weak dependence of the mobility on carrier concentration for both electrons and holes. In particular we predict mobilities ranging from 5 to 
12~cm$^{2}$/Vs for holes and from 2.5 to 10~cm$^{2}$/Vs for electrons. These values are in excellent agreement with the most recent 
measurements~\cite{ADMA:ADMA201305172}. 

When one compares the different phases it appears clear that the cubic structure presents the highest 
mobility for holes and for electrons in the high doping regime. 
Then for holes the mobility of all the other phases is essentially the same, while for electrons there
is more spread with the \textit{tetra2} and \textit{tetra1} displaying the highest and lowest values,
respectively. This essentially follows the size of the relative effective masses, as shown in Tab.~\ref{tab:1}. 
It is however important to remark that at room temperature the molecules in the tetragonal phase 
are almost free to rotate, so that one expects the actual measured mobility to compare with an average of 
those of \textit{tetra1} and \textit{tetra2}.

In light of our results, we suggest that the spectacular light-harvesting performances of perovskite solar cells should not be ascribed to large 
charge carrier mobilities, as previously speculated~\cite{Stoumpos:2013}. Although the values that we have found are to be considered large 
for solution processed materials, they are not 
substantial in absolute terms, in particular if compared to those of conventional semiconductors like Si or Ge, which are three orders of magnitude 
larger. As such, the exceptionally long carrier lifetime of hybrid perovskites should be considered as the principal source of their high efficiencies.

%
A still outstanding issue is whether or not mixing Cl and I (Cl doping) has an impact on the charge 
transport of hybrid perovskites. Indeed, it has been argued that Cl doping would improve the charge transport within the perovskite layer
~\cite{doi:10.1021.cm402919x}, giving rise to better performance in solar cell devices. Other works have shown that mixed halide perovskites 
have a bimolecular recombination rate an order of magnitude lower than that of their tri-iodide counterpart.\cite{ADMA:ADMA201305172}.
In order to partially solve this issue we have calculated the average mobilities of CH$_{3}$NH$_{3}$PbI$_{3-x}$Cl$_{x}$ for $x=1,2$ 
[see lower panel of Fig.~(\ref{fig:4})]. The calculations have been performed here for the tetragonal structure.
Our results show that the mobility assumes values comprised between 2-5~cm$^{2}$/Vs, i.e. it lies within the same order of magnitude than 
CH$_{3}$NH$_{3}$PbI$_{3}$, regardless of the Cl doping level. Clearly our results represent only an upper bound in this case, since impurity
scattering may lower down the relaxation time and hence the mobility. 
In fact, we neglect here the effect arising from 
the spontaneous disorder of alloys, therefore we consider only the crystalline component of the mobility.
Nevertheless, our estimate suggests that the improvement in the device 
efficiency has a different origin, and we endorse the idea that Cl doping does not have much detrimental effects on the transport but may help 
the crystalline growth of the material. 

\begin{figure}[h!]
\includegraphics[width=0.8\columnwidth]{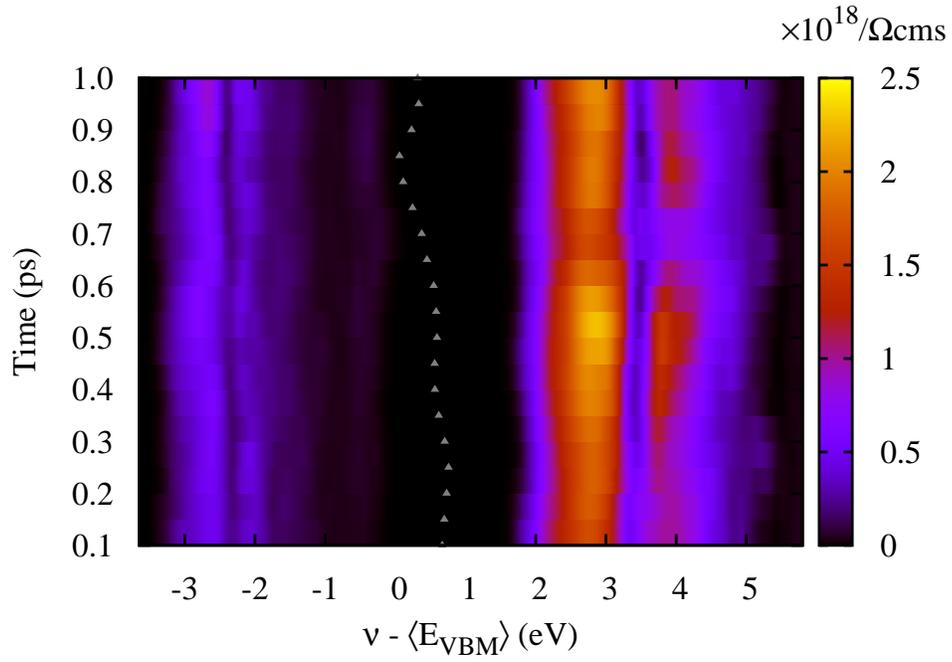} 
\caption{\label{fig:5} Heat map of the charge carrier conductivity $\sigma/\tau$ of CH$_{3}$NH$_{3}$PbI$_{3}$ 
                       as a function of the chemical potential ($\nu$) and the MD time elapsed after an equilibration  phase of 5 ps.
                       The grey triangles represent the total polarization of the molecules contained in the supercell.}
\end{figure}

Finally, we consider an additional aspect. It is understood that dynamic effects play a key role in materials containing organic and 
inorganic sublattices. As mentioned before, at room temperature the organic cations are free to rotate, and recent studies suggest that
the typical time scale for the CH$_{3}$NH$_{3}$ rotation is of the order of 1~ps~\cite{C4CP00569D}. We are thus interested in determining 
the impact of the lattice dynamics on the transport properties of this class of materials. To this end, we have performed Born-Oppenheimer 
molecular dynamics (MD) at 300~K with a cubic 2$\times$ 2$\times$2 supercell of CH$_{3}$NH$_{3}$PbI$_{3}$. The temperature is kept 
constant by the use of the Bussi-Donadio-Parrinello thermostat~\cite{bdp_thermostat}. The system is thermalized with a 5~ps run, and the 
dynamics of the following 1 ~ps is subsequently analyzed. In this time interval, the electronic bandgap is found to have an average value of 
$1.75$~eV and a mean square deviation of $80$~meV. We have sampled 20 equally spaced frames of the last $1$~ps, and computed the 
Boltzmann transport conductivity ($\sigma/\tau$) as previously for the selected configurations. Fig.~\ref{fig:5} displays the heat plot of the 
conductivity as a function of the chemical potential $\nu$ and the time during the last ps of the MD trajectory. By inspecting the trajectory, we 
can also confirm that the cation orientation fluctuates, i.e. that the MA molecule rotates within that time interval. This is supported by the variation 
of the total cation polarization, which is depicted with triangles in the figure. Such quantity is bound between 0 and 1, and clearly spans almost 
all the range within 1~ps. The bandgap variations, due to the molecular rotation, are clearly observed. However, apart from this minor effect, the 
conductivity remains independent of the thermal configurational fluctuations, especially close to the band edges. More noticeable differences 
appear at $\sim 2$~eV above the CBM, which however is not relevant to transport in realistic conditions. This analysis confirm that our results are 
robust against dynamical effects caused by  thermal fluctuations.

As a final consideration, we want to emphasize that our analysis can be taken as a tool for extracting the relaxation time. Indeed the good 
agreement between our calculations and the most recent experiments sets the relaxation time in the ps range. This is the typical time scale 
of the CH$_{3}$NH$_{3}$ rotation at room temperature.
This suggests that the mobility is probably limited by phonon scattering, in particular from those soft vibrations involving the interplay between 
the organic/inorganic sublattices, which is responsible for the carrier relaxation. This work motivates further study on the theoretical determination of 
$\tau$ accounting for the electron-phonon scattering and for the other sources of scattering like impurities and defects.

In summary we have evaluated the charge carrier mobilities of hybrid Pb-halide perovskites, by using a combination of rigorous density functional
theory and semiclassical Boltzmann transport in the constant relaxation time approximation. Our results show that the calculated mobilities are in
the experimental range, once the relaxation time is taken to be 1~ps. The particular crystal structure and the possibility of Cl doping do not 
affect the above results significantly.

\section*{Acknowledgments}
This work is sponsored by the European Research Council, {\sc Quest} project, (CM and SS) and by
the Qatar Environment and Energy Research Institute (FE). Computational resources 
have been provided by the supercomputer facilities at the Trinity Center for High Performance 
Computing, at ICHEC (project tcphy038b) and the research computing at Texas A\&M University at Qatar. 

\section*{Author Contribution}
CM performed the DFT and BoltzTrap calculations. CM, FEM and SS discussed the results and contributed
to the preparation of the manuscript.

\section*{Additional Information}
\textbf{Competing financial interests:} The authors declare no competing financial interests.

\begin{table}[h!]
  \caption{Effective masses calculated with and without SOC for holes 
           (m$_{h}$) and electrons (m$_{e}$) calculated by parabolic fitting 
           along the directions $\Gamma$(0,0,0) $\rightarrow$
           R (0,0,0.5) and  $\Gamma\rightarrow$ Z (0.5,0.5,0). Values are
           given relative to the electron mass.}
  \label{tab:1}
\begin{tabular}{ccccc}
\hline
 & \multicolumn{2}{c}{\textit{tetra1}} &\multicolumn{2}{c}{\textit{tetra2}} \\
\cline{2-5} 
  & m$_{h}$ & m$_{e}$ & m$_{h}$ & m$_{e}$ \\
  \hline
  & \multicolumn{4}{c}{w/o SOC} \\
  $\Gamma \longrightarrow$ R & 0.12 & 0.50 & 0.09 & 0.03 \\
  $\Gamma \longrightarrow$ Z  & 0.11 & 0.12 & 0.11 & 0.36 \\
    & \multicolumn{4}{c}{SOC} \\
  $\Gamma \longrightarrow$ R & 0.19 & 0.21 & 0.10 & 0.07 \\
  $\Gamma \longrightarrow$ Z  & 0.13 & 0.10 & 0.15 & 0.13 \\
\hline
\end{tabular}
\end{table}


\begin{thebibliography}{10}
\expandafter\ifx\csname url\endcsname\relax
  \def\url#1{\texttt{#1}}\fi
\expandafter\ifx\csname urlprefix\endcsname\relax\def\urlprefix{URL }\fi
\providecommand{\bibinfo}[2]{#2}
\providecommand{\eprint}[2][]{\url{#2}}

\bibitem{Green2014}
\bibinfo{author}{Green, M.~A.}, \bibinfo{author}{Ho-Baillie, A.} \&
  \bibinfo{author}{Snaith, H.~J.}
\newblock \bibinfo{title}{The emergence of perovskite solar cells}.
\newblock \emph{\bibinfo{journal}{Nat. Photonics}}
  \textbf{\bibinfo{volume}{8}}, \bibinfo{pages}{506--514}
  (\bibinfo{year}{2014}).

\bibitem{Burschka:2013}
\bibinfo{author}{Burschka, J.} \emph{et~al.}
\newblock \bibinfo{title}{Sequential deposition as a route to high-performance
  perovskite-sensitized solar cells}.
\newblock \emph{\bibinfo{journal}{Nature}} \textbf{\bibinfo{volume}{499}},
  \bibinfo{pages}{316--319} (\bibinfo{year}{2013}).

\bibitem{Liu:2013}
\bibinfo{author}{Liu, M.}, \bibinfo{author}{Johnston, M.~B.} \&
  \bibinfo{author}{Snaith, H.~J.}
\newblock \bibinfo{title}{Efficient planar heterojunction perovskite solar
  cells by vapour deposition}.
\newblock \emph{\bibinfo{journal}{Nature}} \textbf{\bibinfo{volume}{501}},
  \bibinfo{pages}{395--398} (\bibinfo{year}{2013}).

\bibitem{Liu_Kelly_2013}
\bibinfo{author}{Liu, D.} \& \bibinfo{author}{Kelly, T.~L.}
\newblock \bibinfo{title}{Perovskite solar cells with a planar heterojunction
  structure prepared using room-temperature solution processing techniques}.
\newblock \emph{\bibinfo{journal}{Nat. Photonics}}
  \textbf{\bibinfo{volume}{8}}, \bibinfo{pages}{133--138}
  (\bibinfo{year}{2013}).

\bibitem{C4EE00942H}
\bibinfo{author}{Gao, P.}, \bibinfo{author}{Gratzel, M.} \&
  \bibinfo{author}{Nazeeruddin, M.~K.}
\newblock \bibinfo{title}{Organohalide lead perovskites for photovoltaic
  applications}.
\newblock \emph{\bibinfo{journal}{Energy Environ. Sci.}}
  \textbf{\bibinfo{volume}{7}}, \bibinfo{pages}{2448--2463}
  (\bibinfo{year}{2014}).

\bibitem{doi:10.1021.jz5005285}
\bibinfo{author}{Deschler, F.}  \emph{et~al.}
\newblock \bibinfo{title}{High photoluminescence efficiency and optically
  pumped lasing in solution-processed mixed halide perovskite semiconductors}.
\newblock \emph{\bibinfo{journal}{J. Phys. Chem. Lett.}}
  \textbf{\bibinfo{volume}{5}}, \bibinfo{pages}{1421--1426}
  (\bibinfo{year}{2014}).

\bibitem{ADMA:ADMA201305172}
\bibinfo{author}{Wehrenfennig, C.}, \bibinfo{author}{Eperon, G.~E.},
  \bibinfo{author}{Johnston, M.~B.}, \bibinfo{author}{Snaith, H.~J.} \&
  \bibinfo{author}{Herz, L.~M.}
\newblock \bibinfo{title}{High charge carrier mobilities and lifetimes in
  organolead trihalide perovskites}.
\newblock \emph{\bibinfo{journal}{Adv. Mater.}} \textbf{\bibinfo{volume}{26}},
  \bibinfo{pages}{1584--1589} (\bibinfo{year}{2014}).

\bibitem{doi:10.1021jz500858a}
\bibinfo{author}{Savenije, T. J.}  \emph{et~al.}
\newblock \bibinfo{title}{Thermally activated exciton dissociation and
  recombination control the carrier dynamics in organometal halide perovskite}.
\newblock \emph{\bibinfo{journal}{J. Phys. Chem. Lett.}}
  \textbf{\bibinfo{volume}{5}}, \bibinfo{pages}{2189--2194}
  (\bibinfo{year}{2014}).

\bibitem{Poglitsch1987}
\bibinfo{author}{Poglitsch, A.} \& \bibinfo{author}{Weber, D.}
\newblock \bibinfo{title}{Dynamic disorder in
  methylammoniumtrihalogenoplumbates (ii) observed by millimeter-wave
  spectroscopy}.
\newblock \emph{\bibinfo{journal}{J. Chem. Phys}}
  \textbf{\bibinfo{volume}{87}}, \bibinfo{pages}{6373} (\bibinfo{year}{1987}).

\bibitem{Perd96}
\bibinfo{author}{Perdew, J.~P.}, \bibinfo{author}{Burke, K.} \&
  \bibinfo{author}{Ernzerhof, M.}
\newblock \bibinfo{title}{Generalized gradient approximation made simple}.
\newblock \emph{\bibinfo{journal}{Phys. Rev. Lett.}}
  \textbf{\bibinfo{volume}{77}}, \bibinfo{pages}{3865--3868}
  (\bibinfo{year}{1996}).

\bibitem{PhysRevLett.102.073005}
\bibinfo{author}{Tkatchenko, A.} \& \bibinfo{author}{Scheffler, M.}
\newblock \bibinfo{title}{Accurate molecular van der Waals interactions from
  ground-state electron density and free-atom reference data}.
\newblock \emph{\bibinfo{journal}{Phys. Rev. Lett.}}
  \textbf{\bibinfo{volume}{102}}, \bibinfo{pages}{073005}
  (\bibinfo{year}{2009}).

\bibitem{HSE}
\bibinfo{author}{Heyd, J.}, \bibinfo{author}{Scuseria, G.~E.} \&
  \bibinfo{author}{Ernzerhof, M.}
\newblock \bibinfo{title}{Hybrid functionals based on a screened Coulomb
  potential}.
\newblock \emph{\bibinfo{journal}{J. Chem. Phys}}
  \textbf{\bibinfo{volume}{118}}, \bibinfo{pages}{8207--8215}
  (\bibinfo{year}{2003}).

\bibitem{HSE06}
\bibinfo{author}{Paier, J.} \emph{et~al.}
\newblock \bibinfo{title}{Screened hybrid density functionals applied to
  solids}.
\newblock \emph{\bibinfo{journal}{J. Chem. Phys}}
  \textbf{\bibinfo{volume}{124}}, \bibinfo{pages}{154709}
  (\bibinfo{year}{2006}).

\bibitem{PhysRevB.13.5188}
\bibinfo{author}{Monkhorst, H.~J.} \& \bibinfo{author}{Pack, J.~D.}
\newblock \bibinfo{title}{Special points for Brillouin-zone integrations}.
\newblock \emph{\bibinfo{journal}{Phys. Rev. B}} \textbf{\bibinfo{volume}{13}},
  \bibinfo{pages}{5188--5192} (\bibinfo{year}{1976}).

\bibitem{numrec97}
\bibinfo{author}{Press, W.~H.}, \bibinfo{author}{Teukolsky, S.~A.},
  \bibinfo{author}{Vetterling, W.~T.} \& \bibinfo{author}{Flannery, B.~P.}
\newblock \emph{\bibinfo{title}{Numerical recipes}}
  (\bibinfo{publisher}{Cambridge University Press}, \bibinfo{year}{1997}),
  \bibinfo{edition}{3} edn.

\bibitem{g.madsen20064}
\bibinfo{author}{Madsen, G.} \& \bibinfo{author}{Singh, D.}
\newblock \bibinfo{title}{Boltztrap. A code for calculating bandstructure
  dependent quantities}.
\newblock \emph{\bibinfo{journal}{Comput. Phys. Commun.}}
  \textbf{\bibinfo{volume}{175}}, \bibinfo{pages}{6771} (\bibinfo{year}{2006}).

\bibitem{C3TA11222E}
\bibinfo{author}{Zou, D.}, \bibinfo{author}{Xie, S.}, \bibinfo{author}{Liu,
  Y.}, \bibinfo{author}{Lin, J.} \& \bibinfo{author}{Li, J.}
\newblock \bibinfo{title}{Electronic structures and thermoelectric properties
  of layered BiCuOCh oxychalcogenides (Ch = S{,} Se and Te): first-principles
  calculations}.
\newblock \emph{\bibinfo{journal}{J. Mater. Chem. A}}
  \textbf{\bibinfo{volume}{1}}, \bibinfo{pages}{8888--8896}
  (\bibinfo{year}{2013}).

\bibitem{Wu201262}
\bibinfo{author}{Wu, W.}, \bibinfo{author}{Wu, K.}, \bibinfo{author}{Ma, Z.} \&
  \bibinfo{author}{Sa, R.}
\newblock \bibinfo{title}{Doping and temperature dependence of thermoelectric
  properties of AgGaTe2: First principles investigations}.
\newblock \emph{\bibinfo{journal}{Chem. Phys. Lett.}}
  \textbf{\bibinfo{volume}{537}}, \bibinfo{pages}{62 -- 64}
  (\bibinfo{year}{2012}).

\bibitem{PhysRevB.83.085204}
\bibinfo{author}{Lee, M.~S.}, \bibinfo{author}{Poudeu, F.~P.} \&
  \bibinfo{author}{Mahanti, S.~D.}
\newblock \bibinfo{title}{Electronic structure and thermoelectric properties of
  Sb-based semiconducting half-heusler compounds}.
\newblock \emph{\bibinfo{journal}{Phys. Rev. B}} \textbf{\bibinfo{volume}{83}},
  \bibinfo{pages}{085204} (\bibinfo{year}{2011}).

\bibitem{PhysRevB.85.125209}
\bibinfo{author}{Parker, D.} \& \bibinfo{author}{Singh, D.~J.}
\newblock \bibinfo{title}{Thermoelectric properties of AgGaTe2 and
  related chalcopyrite structure materials}.
\newblock \emph{\bibinfo{journal}{Phys. Rev. B}} \textbf{\bibinfo{volume}{85}},
  \bibinfo{pages}{125209} (\bibinfo{year}{2012}).

\bibitem{Stoumpos:2013}
\bibinfo{author}{Stoumpos, C.~C.}, \bibinfo{author}{Malliakas, C.~D.} \&
  \bibinfo{author}{Kanatzidis, M.~G.}
\newblock \bibinfo{title}{Semiconducting tin and lead iodide perovskites with
  organic cations: Phase transitions, high mobilities, and near-infrared
  photoluminescent properties}.
\newblock \emph{\bibinfo{journal}{Inorg. Chem.}} \textbf{\bibinfo{volume}{52}},
  \bibinfo{pages}{9019--9038} (\bibinfo{year}{2013}).

\bibitem{carlo2014}
\bibinfo{author}{Motta, C.} \emph{et~al.}
\newblock \bibinfo{title}{Revealing the role of organic cations in hybrid halide perovskites 
CH$_{3}$NH$_{3}$PbI$_{3}$}  \newblock \emph{\bibinfo{journal}{Nat. Commun.}} 
\newblock \bibinfo{note}{forthcoming} (\bibinfo{year}{2015}).


\bibitem{Umari:2014}
\bibinfo{author}{Umari, P.}, \bibinfo{author}{Mosconi, E.} \&
  \bibinfo{author}{De~Angelis, F.}
\newblock \bibinfo{title}{Relativistic GW calculations on CH3NH3PbI3 and
  CH3NH3SnI3 perovskites for solar cell applications}.
\newblock \emph{\bibinfo{journal}{Sci. Rep.}} \textbf{\bibinfo{volume}{4}}, \bibinfo{pages}{4467}
  (\bibinfo{year}{2014}).

\bibitem{Giorgi:2013}
\bibinfo{author}{Giorgi, G.}, \bibinfo{author}{Fujisawa, J.~I.},
  \bibinfo{author}{Segawa, H.} \& \bibinfo{author}{Yamashita, K.}
\newblock \bibinfo{title}{Small photocarrier effective masses featuring
  ambipolar transport in methylammonium lead iodide perovskite: A density
  functional analysis}.
\newblock \emph{\bibinfo{journal}{J. Phys. Chem. Lett.}}
  \textbf{\bibinfo{volume}{4}}, \bibinfo{pages}{4213--4216}
  (\bibinfo{year}{2013}).

\bibitem{doi:10.1021.cm402919x}
\bibinfo{author}{Colella, S.} \emph{et~al.}
\newblock \bibinfo{title}{MAPbI3-xClx mixed halide perovskite for hybrid solar
  cells: The role of chloride as dopant on the transport and structural
  properties}.
\newblock \emph{\bibinfo{journal}{Chem. Mater.}} \textbf{\bibinfo{volume}{25}},
  \bibinfo{pages}{4613--4618} (\bibinfo{year}{2013}).

\bibitem{C4CP00569D}
\bibinfo{author}{Mosconi, E.}, \bibinfo{author}{Quarti, C.},
  \bibinfo{author}{Ivanovska, T.}, \bibinfo{author}{Ruani, G.} \&
  \bibinfo{author}{De~Angelis, F.}
\newblock \bibinfo{title}{Structural and electronic properties of organo-halide
  lead perovskites: a combined IR-spectroscopy and ab initio molecular dynamics
  investigation}.
\newblock \emph{\bibinfo{journal}{Phys. Chem. Chem. Phys.}} \textbf{\bibinfo{volume}{16}},
  \bibinfo{pages}{16137--16144} (\bibinfo{year}{2014}).

\bibitem{bdp_thermostat}
\bibinfo{author}{Bussi, G.}, \bibinfo{author}{Donadio, D.} \&
  \bibinfo{author}{Parrinello, M.}
\newblock \bibinfo{title}{Canonical sampling through velocity rescaling}.
\newblock \emph{\bibinfo{journal}{Journal of Chemical Physics}}
  \textbf{\bibinfo{volume}{126}}, \bibinfo{pages}{014101}  (\bibinfo{year}{2007}).

\end{thebibliography}
\end{document}